\documentclass{PoS}
\usepackage{wrapfig}
\usepackage{url}
\usepackage{algorithm}
\usepackage{algorithmic}
\makeatletter
\newcommand{\figcaption}[1]{\def\@captype{figure}\caption{#1}}
\newcommand{\tblcaption}[1]{\def\@captype{table}\caption{#1}}

\newcommand{\SPARC}{SPARC64$^{\mathrm{TM}}$ VIIIfx }

\makeatletter
\newcommand{\rmnum}[1]{\romannumeral #1}
\newcommand{\Rmnum}[1]{\expandafter\@slowromancap\romannumeral #1@}
\makeatother

\title{Multi-block/multi-core SSOR preconditioner for the QCD quark solver for K computer}

\ShortTitle{Multi-block/multi-core SSOR preconditioner for the QCD quark solver for K computer}

\author{
T.~Boku$^{a}$,
\speaker{K.-I. Ishikawa}$^{b,c}$,
Y.~Kuramashi$^{a,c,f}$,
K.~Minami$^{c}$,
Y.~Nakamura$^{c}$,
F.~Shoji$^{c}$,
D.~Takahashi$^{a}$,
M.~Terai$^{c}$,
A.~Ukawa$^{a}$,
T.~Yoshi\'e$^{a,f}$\\
$^{a}$Center for Computational Sciences, University of Tsukuba, Tsukuba, Ibaraki 305-8577, Japan\\
$^{b}$Graduate School of Science, Hiroshima University, Higashi-Hiroshima, Hiroshima 739-8526, Japan\\
$^{c}$RIKEN Advanced Institute for Computational Science, Kobe, Hyogo 650-0047, Japan\\
$^{f}$Faculty of Pure and Applied Sciences, University of Tsukuba, Tsukuba, Ibaraki 305-8571, Japan\\
E-mail: \email{ishikawa@theo.phys.sci.hiroshima-u.ac.jp}}

\abstract{
We study the algorithmic optimization and performance tuning of 
the Lattice QCD clover-fermion solver for the K computer.
We implement the L\"{u}scher's SAP preconditioner with sub-blocking in which the lattice block 
in a node is further divided to several sub-blocks to extract enough parallelism for 
the 8-core CPU \SPARC of the K computer.
To achieve a better convergence property we use 
the symmetric successive over-relaxation (SSOR) iteration with
{\it locally-lexicographical} ordering for the sub-blocks in obtaining the block inverse.
The SAP preconditioner is included in the single precision BiCGStab solver of 
the nested BiCGStab solver.
The single precision part of the computational kernel are solely written with 
the SIMD oriented intrinsics to achieve the best performance of the \SPARC on the K computer.
We benchmark the single precision BiCGStab solver on the three lattice sizes: 
$12^3\times 24$, $24^3\times 48$ and $48^3\times 96$, 
with fixing the local lattice size in a node at $6^3\times 12$. 
We observe an ideal weak-scaling performance from $16$ nodes to $4096$ nodes.
The performance of a computational kernel exceeds 50\% efficiency, and 
the single precision BiCGstab has $\sim$26\% susutained efficiency.
}

\FullConference{The 30th International Symposium on Lattice Field Theory\\
		 June 24 -- 29,  2012\\
		 Cairns, Australia}

\begin{document}
\section{Lattice QCD on the K computer}

The K computer has been developed by RIKEN and Fujitsu since 2006 as the national 
leadership computer of Japan to promote science and technology~\cite{KcomputerAICS}.
The construction has been finished on July 2012 and the system is now provided to 
the strategic programs of the High Performance Infrastructure (HPCI) and 
to the research users those who apply the computer resource through the HPCI~\cite{HPCI}.
The strategic programs consist of five fields:
(\rmnum{1}) Predictable life science, healthcare and drug discovery foundation, 
(\rmnum{2}) New Materials and Energy Creation,
(\rmnum{3}) Projection of Planet Earth Variations for Mitigating Natural Disasters,
(\rmnum{4}) Next-generation manufacturing technology, and
(\rmnum{5}) The origin of matter and the universe. 
Lattice QCD is contained in the fifth strategic program~\cite{HPCISPFFive}.

The K computer consists of over 80,000 computational nodes connected by 
the so called ``Tofu'' network. 
The Tofu network topology is six dimensional topology with 3D-mesh times 3D-torus shape, 
with which robustness against the node failure and high node flexibility 
and availability are ensured.
Each node has a single CPU chip called ``\SPARC''. 
Equipping 8 cores with SIMD enabled 256 registers and 6MB shared L2 cache, 
the CPU can achieve a high efficiency for scientific applications
(for the detailed system structure see~\cite{FSTJ}).
Lattice QCD is one of the suitable applications for this kind of 
massively parallel system architecture.

In this paper we present the algorithmic and performance tuning of the $O(a)$ improved 
Wilson quark solver for the K computer. 
We focus on the solver algorithm preconditioned by the domain-decomposed 
Schwartz alternating procedure (SAP) with mixed precision~\cite{LuscherSAPHMC}.
In the next section we briefly introduce the SAP preconditioner and
the nested BiCGStab algorithm~\cite{NestedBiCGStab}.
The optimization of the algorithm and tuning methods for the computational 
kernels of the SAP are presented in section~\ref{seq:Tunning}.
We give the performance benchmark results in section \ref{seq:Results} and summarize 
the paper in the last section.

\section{Nested BiCGStab with L\"{u}scher's SAP preconditioner}

Our target problem is solving the following linear equation:
\begin{equation}
    D x = b,
\label{eq:LinEQ}
\end{equation}
where
$D$ is the clover term preconditioned $O(a)$-improved Wilson-Dirac operator:
\begin{eqnarray}
    D^{a,b}_{\alpha,\beta}(n,m) &=& \delta^{a,b}\delta_{\alpha,\beta}\delta(n,m)
- \kappa F^{a,c}_{\alpha,\gamma}(n) \sum_{\mu=1}^{4}
\left[ 
 ( 1-\gamma_{\mu})_{\gamma,\beta} (U_{\mu}(n))^{c,b}     \delta(n+\hat{\mu},m)\right.\nonumber\\
&&\left.\hspace{12em}
+( 1+\gamma_{\mu})_{\gamma,\beta} ((U_{\mu}(m))^{b,c})^* \delta(n-\hat{\mu},m)
\right].
\end{eqnarray}
Where $F(n)$ is the inverse clover term $(1-(c_{\mathrm{SW}}\kappa/2)\sigma_{\mu\nu}F_{\mu\nu}(n))^{-1}$, 
$(n,m)$ are lattice site, $a,b$ color, $\alpha,\beta$ are spin indexes.

L\"{u}scher has introduced the Schwartz alternating procedure (SAP)
to further preconditioning the Eq.~(\ref{eq:LinEQ})~\cite{LuscherSAPHMC}. 
By dividing the whole lattice into two colored blocks in checkerboard manner,
Eq.~(\ref{eq:LinEQ}) can be rewritten 
in the following $2 \times 2$ block form:
\begin{equation}
    \left(
        \begin{array}{cc}
 D_{EE} & D_{EO}\\
 D_{OE} & D_{OO}
        \end{array}\right)
\left(
        \begin{array}{c}
            x_{E}\\ x_{O}
        \end{array}\right)=
\left(
        \begin{array}{c}
            b_{E}\\ b_{O}
        \end{array}\right),
\end{equation}
where $D_{EE}$ ($D_{OO}$) is a block restricted operator in even-domain (odd-domain),
while $D_{EO}$ ($D_{OE}$) contains hopping operations from odd-domain to even-domain 
(and vice versa). 
The SAP practitioner $M_{\mathrm{SAP}}$ is introduced as
\begin{equation}
 M_{\mathrm{SAP}} = K\sum_{j=0}^{N_{\mathrm{SAP}}-1} (1-DK)^{j},\quad\quad \mbox{with} \quad\quad
K= \left(
    \begin{array}{cc}
        A_{EE} & 0 \\
      -A_{OO}D_{OE}A_{EO} & A_{OO} \\
    \end{array}
\right),
\end{equation}
where $A_{EE}$ ($A_{OO}$) can be any approximation for $(D_{EE})^{-1}$ ($(D_{OO})^{-1}$). 
When $A_{EE}$ and $A_{OO}$ are exact, $DK$ becomes block triangular and is expected to be well 
preconditioned.
In the case $|DK|<1$, $M_{\mathrm{SAP}}$ converges to $D^{-1}$ 
when $N_{\mathrm{SAP}}\rightarrow \infty$.

The nested BiCGStab solver is designed to be flexible against the preconditioner changing 
iteration by iteration~\cite{NestedBiCGStab} using an inner-outer strategy.
The outer BiCGStab contains the inner BiCGStab solver as the flexible preconditioner.
The flexibility ensures the double precision accuracy of the solution vector even if 
we use the single precision for the inner BiCGStab solver~\cite{mixedprec}. 
We apply the SAP preconditioner $M_{\mathrm{SAP}}$ to the inner single precision BiCGStab 
solver.
The use of single precision has a merit as it requires less system resources than 
double precision.
We target the single precision part of the solver as the tuning part for the K computer.
In the following section we focus on the tuning of the single precision $DK$ of $M_{\mathrm{SAP}}$.

\section{Performance tuning for the K computer}
\label{seq:Tunning}

\paragraph{Inverse of block operator and OpenMP threading}

The approximate inverse of the block operator $A_{EE}$ ($A_{OO}$) are the important part of the SAP.
The exactness is not required for the SAP, however, the better approximation 
with less computational cost is preferred for $A_{EE}$ ($A_{OO}$). 
The even-odd site preconditioning has been used in~\cite{LuscherSAPHMC} and
the SSOR preconditioning has been used in~\cite{PACSCSLDDHMC}. 
The latter has a better performance than the former at the same computational cost.
The SSOR preconditioning is derived by decomposing the original operator into 
the sum of an upper and a lower triangular matrices. 
The preconditioning is achieved by solving the upper and lower triangular parts
through forward and backward substitutions.
The decomposition and the efficiency of the SSOR depend on the site ordering.
It is observed that the SSOR with natural ordering have a better performance~\cite{PACSCSLDDHMC}.
However the SSOR with natural ordering is not suitable for the multi-core CPU architecture
because the natural ordering has a global data recurrence pattern and less parallelism in 
the forward and backward substitutions.
To extract 8 core parallelism for the K computer, 
we further divide the block into 16 sub-blocks via 
the {\it locally-lexicographical} ordering ({\it ll}-ordering) described in~\cite{LLSSOR}.

\begin{wrapfigure}{c}{0.59\hsize}
\vspace*{-1em}
    \centering
    \includegraphics[scale=0.47,trim=20 20 20 10]{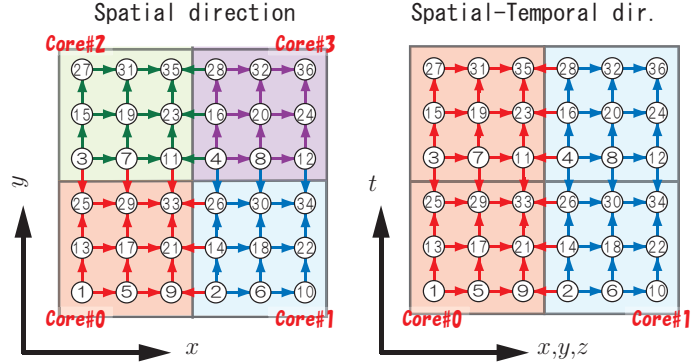}
    \caption{Site ordering in a CPU. 
    This is an example with a $6^4$ block in a node. 
    The numbering of sites is limited in the 2-D plane for simplicity.}
\vspace*{-1em}
    \label{fig:BSSOR}
\end{wrapfigure}

Figure~\ref{fig:BSSOR} shows an example of $6^4$ lattice in a node. 
The block in a node is divided into $2^4$ sub-blocks.
Each single core contains two sub-blocks (two sub-blocks with size $3^4$ adjacent in temporal direction). 
The numbering on the sites is the ordering according to the {\it ll}-ordering.
The arrows on the links represent the data recurrence direction.
The most of the recurrence are limited in each sub-block and 
there are little data reference on the surface of sub-blocks.
Based on this ordering we wrote the forward and backward solver
to construct $A_{EE}$ ($A_{OO}$) with the SSOR.  
The parallelization in a node is achieved by explicit OpenMP threading.
8 threads are invoked and two sub-blocks are assigned to each OpenMP thread.
The spatial division is mapped to OpenMP 8 threads, and the temporal division is dedicated 
to the loop unrolling in a single thread.
To resolve the recurrence dependency among threads, we  carefully
insert explicit {\bf omp barrier}'s at the sites that require data on other threads.

The sub-block size affects the performance of the SAP. 
It has been reported that the size of $2^4$ for block still 
has a gain against the even-odd site ordering on a larger size 
lattice~\cite{LLSSOR}. 
Although we apply the {\it ll}-ordering to a small lattice (corresponds to a block in a node),
we observed that the use of the SSOR with sub-blocking via {\it ll}-ordering
for $A_{EE}$ still has a gain against the even-odd site ordering.

\paragraph{SIMDzation}

The \SPARC CPU has 256 double precision (64bit) FP registers per core.
Each core can issue two independent fused multiply add (FMAD) on paired registers (SIMD) 
in a cycle resulting in 8 floating operations per cycle.
To fully make use of these core infrastructure 
we have to extract enough independent data-computation stream
in the kernel code $D_{EE}$ and $A_{EE}$.

The SIMD FMAD operation can be explicitly invoked using 
intrinsic functions defined in the C/C++ language provided by Fujitsu.
Although our entire program codes were developed in the Fortran90 at first,
we totally write the single precision inner BiCGStab solver using the SIMD intrinsic 
functions in the C language.
The details of the intrinsics are not available publicly, while these have one-to-one 
correspondence to the mnemonic of the \SPARC~\cite{SPARCACE}.
The linkage between the Fortran90 and the C codes is realised through the \verb+ISO_C_BINDING+ 
feature of the Fortran 2003 which is partly included in the Fujitsu Fortran.
We use double precision intrinsic functions for the SIMD computation because
the single precision FLOPS performance is identical to that in double precision
in this CPU architecture.
The data are converted from single precision to double precision (and vice verse) 
at the data loading (storing) from (to) memory.
We employ the $SU(3)$ reconstruction method in the kernel code where  
the link variables are stored in compressed form (keeps only two-columns of the $SU(3)$ matrix) 
and the third column is reconstructed using the unitarity condition on the fly.

\begin{wraptable}{r}{0.45\hsize}
\vspace*{-1em}
    \centering
    \begin{tabular}{c|ccc}\hline
 \raisebox{0pt}[23pt][0pt]{\shortstack{$SU(3)$\\ reconst.}} & 
 \shortstack{Flop\\ count}  &
 \shortstack{Load/Store\\ count (real)} &
 \raisebox{0.6em}[0pt][0pt]{Real/Flop} \\\hline
    Yes  & 2232  & 372        & 0.1667  \\
     No  & 1896  & 420        & 0.2215  \\\hline
    \end{tabular}
    \caption{Flop count and load/store count for the clover hopping matrix per site.}
\vspace*{-1em}
    \label{tab:flopc}
\end{wraptable}

To further improve the kernel performance we unroll the temporal direction 
loop, which is the innermost site loop, by three times for $D_{EE}$. 
For the forward/backward solver in $A_{EE}$
the two independent sub-blocking in temporal direction is used as unrolling. 
This is based on the technique of the register blocking which hide the data load latency.
The unrolling also contains branch elimination by condition merging.
This loop unrolling with huge loop body is possible thanks to the many registers 
of the \SPARC.

In table~\ref{tab:flopc} 
we summarize the total flop and load/store count per site for the hopping matrix of $D$. 
These numbers are estimated for bulk sites in a block (not for the surface sites).
We use the Dirac representation for $\gamma_{\mu}$ ($\gamma_4$ is diagonal)
in the spin projection, and use the chiral representation for the inverse clover 
term $F(n)$. When we multiply $F(n)$, it is converted to the Dirac representation on the fly. 
The required byte/flop is 0.667 with the $SU(3)$ reconstruction method in single precision.
The theoretical system byte/flop is 64 [GByte/s]/128 [GFlops] = 0.5. 
Thus our computation is still limited by the memory bandwidth.
The maximum theoretical performance is estimated to be $\sim$ 96 GFlops 
from the system memory bandwidth of 64 GByte/s. 
Excluding the redundant flop count caused by the $SU(3)$ reconstruction, 
the effective performance is $\sim$ 82 GFlops, which is still better than that without 
the $SU(3)$ reconstruction ($\sim 72$ GFlops).
This performance number could be reduced or enhanced by the cache system, site 
loop control, conditional branch for site location, thread controlling etc.
We test and benchmark the $D_{EE}$ and $A_{EE}$ kernels.

\begin{wraptable}{r}{0.49\hsize}
\vspace*{-2em}
\begin{minipage}[H]{\hsize}
\begin{algorithm}[H]{\small
  \caption{$w=DKv$ computation with communication hiding. 
           $f_E$ and $f_O$ are working vectors.}
  \label{alg:MatMulDK}
  \begin{algorithmic}[1]
        \STATE $x_{E} = A_{EE} v_{E}$
        \STATE {\bf MPI\_Isend} and {\bf MPI\_Irecv} for $w_O = D_{OE} x_{E}$.
        \STATE $w_E = D_{EE} x_E$
        \STATE {\bf MPI\_Wait} for $w_O = D_{OE} x_{E}$.
        \STATE $f_O = v_O - w_O$
        \STATE $x_{O} = A_{OO} f_O$
        \STATE {\bf MPI\_Isend} and {\bf MPI\_Irecv} for $f_E = D_{EO} x_{O}$.
        \STATE $f_O = D_{OO} x_O$
        \STATE $w_O = w_O + f_O$
        \STATE {\bf MPI\_Wait} for $f_E = D_{EO} x_{O}$.
        \STATE $w_E = w_E + f_E$
  \end{algorithmic}}
\end{algorithm}
\vspace*{-2em}
\end{minipage}
\end{wraptable}

\paragraph{Communication hiding}

The internode communication in the SAP kernel $DK$ is arranged in
the $D_{EO}$ and $D_{OE}$ kernels.
The structure of $D_{EO}$ ($D_{OE}$) is basically organized as follows:
(\rmnum{1}) spin projection, link $U^{\dag}_{\mu}(m)$ multiplication and data packing, 
(\rmnum{2}) sending data,
(\rmnum{3}) receiving data,
(\rmnum{4}) link $U_{\mu}(n)$ multiplication, spin reconstruction and accumulation.
Other computation is possible during the step (\rmnum{2}) and (\rmnum{3}).

We organize the SAP kernel $DK$ to hide the communication time of $D_{EO}$ ($D_{OE}$)
behind the computation of $D_{EE}$ ($D_{OO}$) as shown in Alg.~\ref{alg:MatMulDK}.
To hide the communication we employ the non-blocking MPIs:
 {\bf MPI\_Isend}, {\bf MPI\_Irecv}, and {\bf MPI\_Wait}. 
The steps (\rmnum{1})-(\rmnum{2}) are done at the lines 2 and 7, and 
the steps (\rmnum{3})-(\rmnum{4}) are at the lines 4 and 10.
We benchmark the communication performance in the weak-scaling test by
comparing the performance of $D_{EE}$ and $DM_{\mathrm{SAP}}$, where 
the latter contains the internode communication while the former does not.

\section{Results}
\label{seq:Results}

We benchmark the single precision BiCGStab with the SAP preconditioner on the K computer.
The lattice sizes benchmarked are $12^3\times 24$, $24^3\times 48$, and $48^3\times 96$.
The block size of the SAP is kept fixed at $6^4$ and the local lattice size in a node is 
$6^3\times 12$.  The sub-block size for the SSOR is thus $3^4$.
The number of nodes used for the benchmark is 16,\  256 and 4096 nodes.
The performance is measured using the profiler provided by the Fujitsu's compiler 
system.

\begin{figure}[t]
\vspace*{-1em}
\begin{minipage}[t]{0.47\hsize}
    \centering
    \includegraphics[scale=0.58,trim= 20 20 20 20]{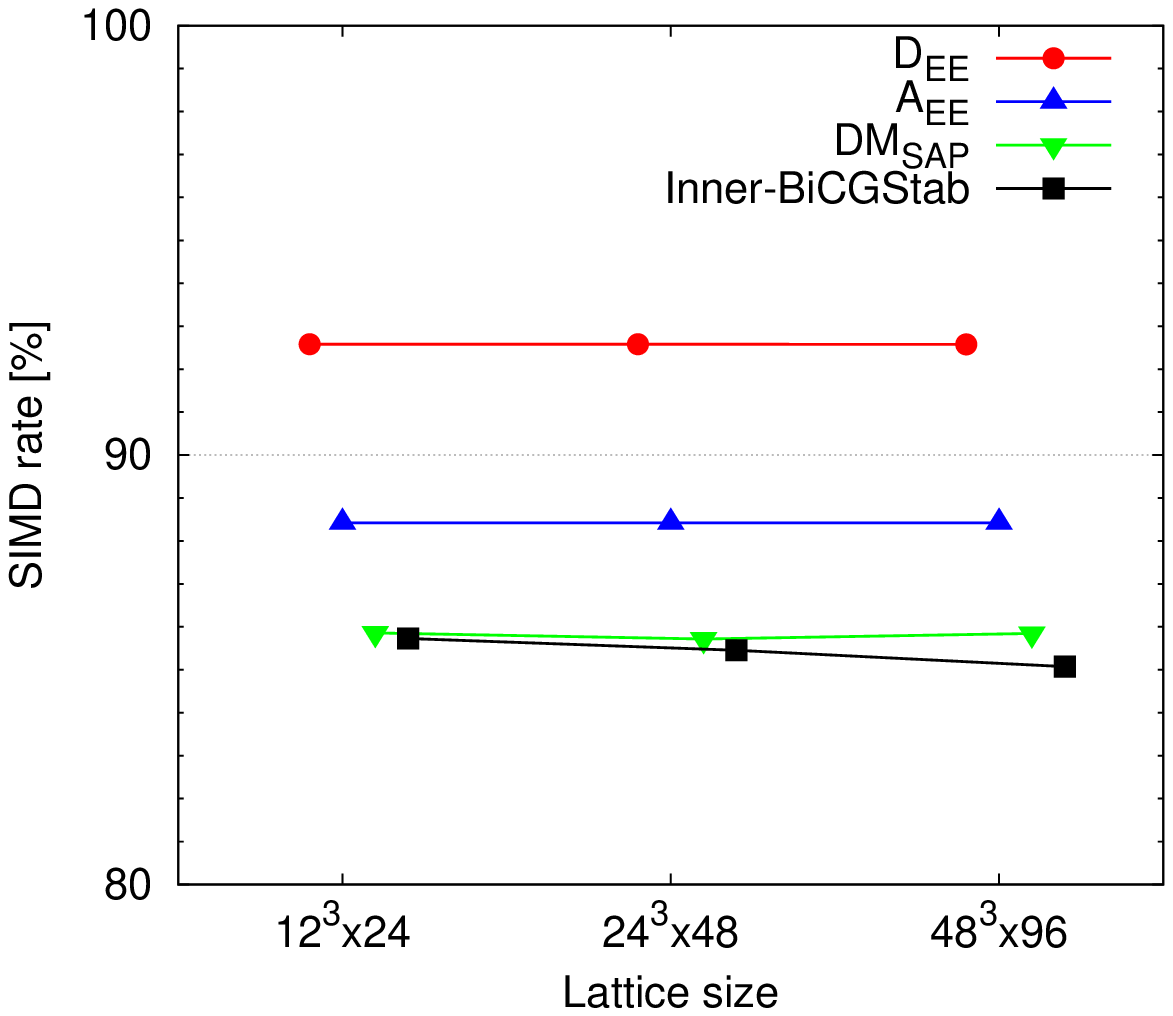}
    \caption{Weak-scaling of the SIMD rate of the single precision kernels in a node.}
    \label{fig:SIMD}
\end{minipage}
\hfill
\begin{minipage}[t]{0.47\hsize}
    \centering
    \includegraphics[scale=0.58,trim= 20 20 20 20]{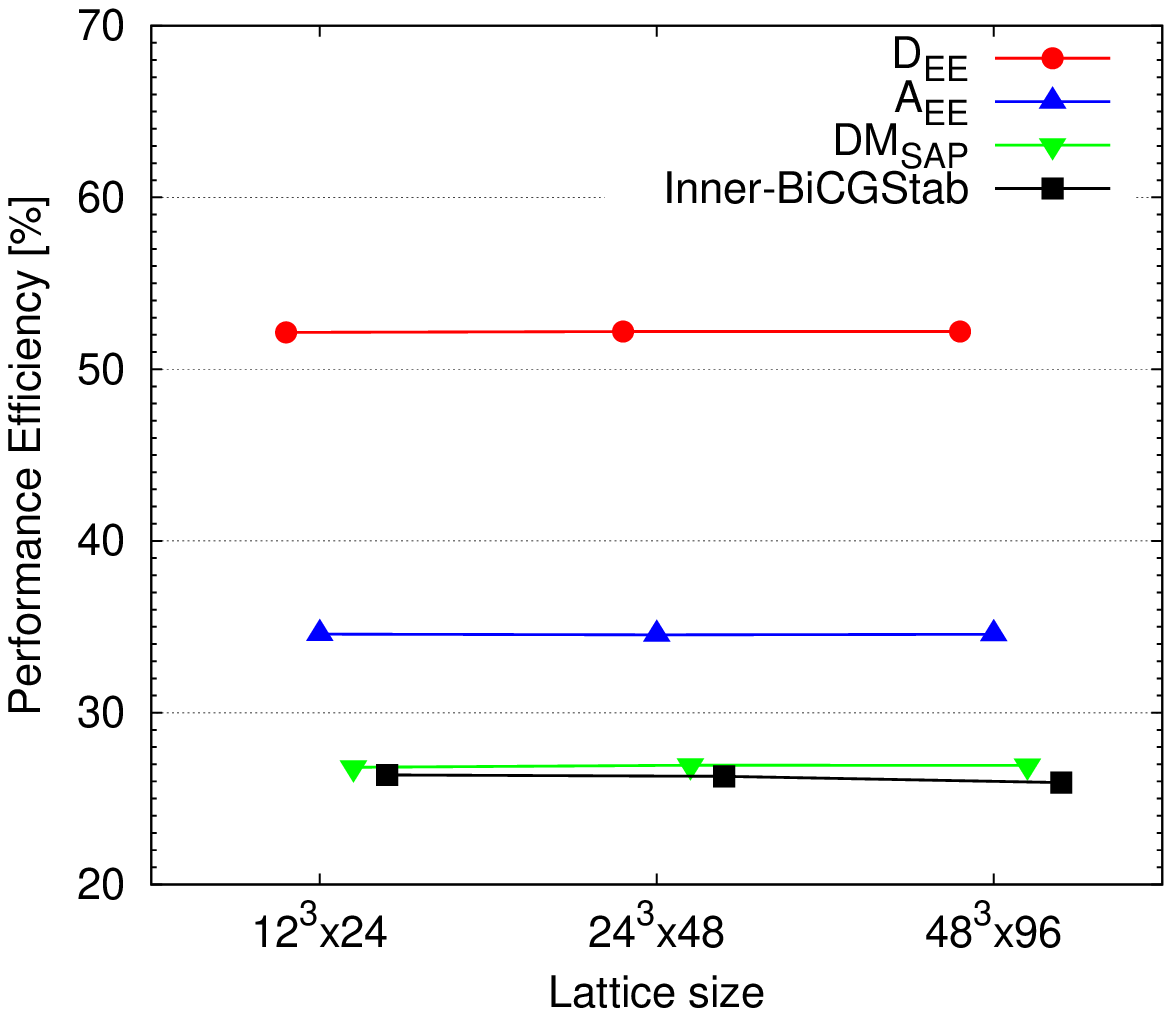}
    \caption{Same as Fig.~{\protect\ref{fig:SIMD}}, but for the performance efficiency in a node.}
    \label{fig:EFF}
\end{minipage}
\vspace*{-1em}
\end{figure}

\begin{wrapfigure}{r}{0.48\hsize}
\vspace*{-0em}
    \centering
    \includegraphics[scale=0.58,trim= 20 20 20 10]{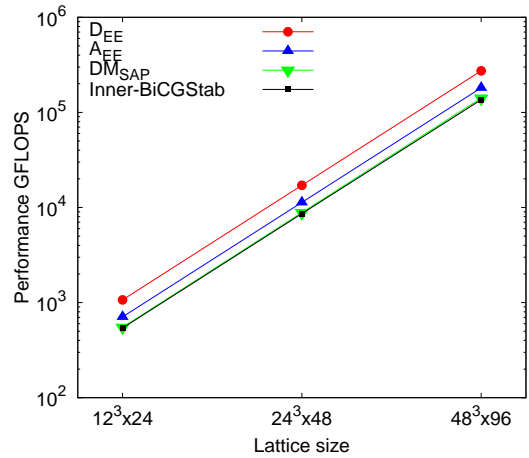}
    \caption{Weak-scaling of the total flops performance.}
    \label{fig:FLOPS}
\vspace*{-.5em}
\end{wrapfigure}

Figure~\ref{fig:SIMD} shows the SIMD rate in the instruction executed in benchmarking runs. 
We achieve 90\% SIMD rate for the kernel $D_{EE}$ and $A_{EE}$ by 
explicitly using the SIMD intrinsic functions.
At the solver level it reduces to 85\% as it 
contains internode communication.
Figs~\ref{fig:EFF} and \ref{fig:FLOPS} represents
 the weak-scaling property of the flops performance.
The efficiency (Fig.~\ref{fig:EFF}) is almost at constant for the kernels
resulting an ideal weak-scaling (Fig.~\ref{fig:FLOPS}) from 16 nodes to 4096 nodes.
\noindent
The efficiency reaches over 50\% for the $D_{EE}$ kernel, while 
the $A_{EE}$ kernel has lower 35\%.

There are two reasons for the lower performance for $A_{EE}$.
One is the load imbalance of the computation among the threads in $A_{EE}$.
For example, the number of red arrows is more than that of purple in 
the left panel of figure~\ref{fig:BSSOR}. This means that 
the core \#0 has much computation than that of the core \#3. 
The second is the less computational density in the loop body compared to that of $D_{EE}$.
As mentioned in previous section the loop of $D_{EE}$ is unrolled by three times,
while $A_{EE}$ is by two with sub-blocking. 
The forward/backward substitution of $A_{EE}$ only has one sided hopping operation for bulk sites.
This also reduces the computational density in the loop body.
The theoretical peak efficiency for $D_{EE}$ is estimated be 75\%($=$96 [GFlops]/128 [GFlops]),
while the observed 50\% efficiency does not reach the theoretical peak efficiency.
A reason for this could be the conditional branch to distinguish the sites 
in the bulk or on the surface of the block. We still need a further investigation 
on the gap between the theoretical peak efficiency and the observed efficiency.

The performance efficiency of the single precision inner solver is at $\sim$26\% in all.
We observe an ideal weak-scaling property as shown in Fig.~\ref{fig:FLOPS}.
Note that the measured performance presented in the figures contains 
redundant floating operations coming from the $SU(3)$ reconstruction and 
the use of FMAD for the spin projection.
We have to roughly multiply 0.8
on the Flops numbers in Fig.~\ref{fig:FLOPS}
to exclude the redundant operations and to obtain the effective performance.

\section{Summary}

We have benchmarked the single-precision BiCGStab solver for the $O(a)$-improved Wilson 
fermion on the K computer. 
The solver code has been successfully optimized and tuned for the system. 
We have applied the SSOR with the {\it ll}-ordered sub-blocking for the approximate block inverse 
of the SAP preconditioner to enhance the parallelism for the multi-core architecture.
Thanks to the science specific architecture of th K computer we could achieve 
the ideal weak-scaling performance and $\sim 26\%$ efficiency for the 
single-precision BiCGStab solver. 
It partly remains unclear the origin of the gap between the theoretical 
system performance and the measured performance. 
We expect a further improvement on the internode communication by using 
the ``Tofu'' specific optimization which is not covered in this study.

\paragraph*{Acknowledgement}
This work has been done in the 
``collaborative research for the performance analysis of large scale simulations
on next-generation supercomputers'' between RIKEN and University of Tsukuba.
We thank the members of the next-generation Technical Computing Unit of Fujitsu 
for giving us the technical advice and support, and for tuning the computational kernels.
Part of the results is obtained by using the K computer at the RIKEN 
Advanced Institute for Computational Science (Proposal numbers hp120108, hp120153, hp120170, hp120281),
T2K-Tsukuba System in Center for Computational Sciences, University of Tsukuba,
and a PC-cluster of HPCI strategic program Field 5. 
This work is supported in part by Grants-in-Aid for Scientific Research from 
the Ministry of Education, Culture, Sports, Science and Technology
(Nos. 22244018, 24540276).


\begin{thebibliography}{99}

\bibitem{KcomputerAICS}
RIKEN Advanced Institute for Computational Science (AICS), 
[\url{http://www.aics.riken.jp/en/}].

\bibitem{HPCI}
High Performance Computing Infrastructure (HPCI), [\url{https://www.hpci-office.jp/index.html}].

\bibitem{HPCISPFFive} 
HPCI Strategic Program Field 5, \emph{The origin of matter and the universe}, [\url{http://www.jicfus.jp/field5/en/}].

\bibitem{FSTJ} FUJITSU SCIENTIFIC \& TECHNICAL JOURNAL (FSTJ), \emph{The K computer},
2012-7 (Vol.48, No.3) [\url{http://www.fujitsu.com/global/news/publications/periodicals/fstj/}].

\bibitem{LuscherSAPHMC}
  M.~L\"{u}scher,
    Comput.\ Phys.\ Commun.\  {\bf 165} (2005) 199
  [arXiv:hep-lat/0409106];
  JHEP {\bf 0305} (2003) 052
  [hep-lat/0304007].

\bibitem{NestedBiCGStab}
 J.A.~Vogel, Appl.\ Math.\ Comput, {\bf 167} (2005) 1004-1025;
 H.~Tadano and T.~Sakurai, LSSC'07, Lec.\ Notes Comput.\ Sci.\ {\bf 4818} (2008) 721.

\bibitem{mixedprec}
A.~Buttari, J.~Dongarra, J.~Kurzak, P.~Luszczek, and S.~ Tomov,
\emph{Using mixed precision for sparse matrix computations to 
enhance the performance while achieving 64-bit accuracy},
ACM Trans.\ Math.\ Soft., {\bf 34} (2008) 1.

\bibitem{PACSCSLDDHMC}
  S.~Aoki {\it et al.}  [PACS-CS Collaboration],
  Phys.\ Rev.\ D {\bf 81} (2010) 074503
  [arXiv:0911.2561 [hep-lat]];
  Phys.\ Rev.\ D {\bf 79} (2009) 034503
  [arXiv:0807.1661 [hep-lat]].

\bibitem{LLSSOR}
  N.~Eicker, W.~Bietenholz, A.~Frommer, T.~Lippert, B.~Medeke and K.~Schilling,
  Nucl.\ Phys.\ Proc.\ Suppl.\  {\bf 73} (1999) 850
  [arXiv:hep-lat/9809038];
  S.~Fischer, A.~Frommer, U.~Glassner, T.~Lippert, G.~Ritzenhofer and K.~Schilling,
  Comput.\ Phys.\ Commun.\  {\bf 98} (1996) 20
  [arXiv:hep-lat/9602019].

\bibitem{SPARCACE}
  Fujitsu Limited, ``SPARC64$^{\mathrm{TM}}$ VIIIfx Extensions'' Version 15, 2010.

\end{thebibliography}
\end{document}